\def\year{2018}\relax
\documentclass[letterpaper]{article} 
\usepackage{aaai19}  
\usepackage{times}  
\usepackage{helvet}  
\usepackage{courier}  
\usepackage{url}  
\usepackage{graphicx}  
\usepackage{booktabs}  
\usepackage{enumitem}
\usepackage{amsmath}
\usepackage{pdfpages} 
\usepackage{float}
\usepackage{placeins}

\begin{document}
%
\title{PhoneMD:\\ Learning to Diagnose Parkinson's Disease from Smartphone Data}
\author{Patrick Schwab  \\
Institute of Robotics and Intelligent Systems\\
ETH Zurich, Switzerland \\
patrick.schwab@hest.ethz.ch  \And Walter Karlen\\
Institute of Robotics and Intelligent Systems\\
ETH Zurich, Switzerland\\
walter.karlen@ieee.org \\
}
\maketitle
\begin{abstract}
Parkinson's disease is a neurodegenerative disease that can affect a person's movement, speech, dexterity, and cognition. Clinicians primarily diagnose Parkinson's disease by performing a clinical assessment of symptoms. However, misdiagnoses are common. One factor that contributes to misdiagnoses is that the symptoms of Parkinson's disease may not be prominent at the time the clinical assessment is performed. Here, we present a machine-learning approach towards distinguishing between people with and without Parkinson's disease using long-term data from smartphone-based walking, voice, tapping and memory tests. We demonstrate that our attentive deep-learning models achieve significant improvements in predictive performance over strong baselines (area under the receiver operating characteristic curve = 0.85) in data from a cohort of 1853 participants. We also show that our models identify meaningful features in the input data. Our results confirm that smartphone data collected over extended periods of time could in the future potentially be used as a digital biomarker for the diagnosis of Parkinson's disease.
\end{abstract} 
\section{Introduction} 
Parkinson's disease (PD) affects more than 6 million people worldwide \cite{vos2016global} and is the second most common neurodegenerative disease after Alzheimer's disease \cite{de2006epidemiology}. The symptoms of PD progressively worsen over time, leading to a stark loss in quality of life \cite{schrag2000does}, and a significant reduction in life expectancy \cite{de2006epidemiology}. While there currently exists no cure for PD, available pharmacological and surgical treatment options are effective at managing the symptoms of PD \cite{goetz2005evidence,connolly2014pharmacological}. Receiving a timely and accurate diagnosis is paramount for patients because access to treatments could improve their quality of life \cite{global2002factors}. Currently, clinicians diagnose PD primarily based on subjective clinical assessments of patients' symptoms \cite{pahwa2010early}. However, research has shown that around 25\% of PD diagnoses are incorrect when compared to results of post-mortem autopsy \cite{pahwa2010early}. Diagnosing PD is difficult because there are other movement disorders that may appear similar to PD, and because symptom severity in PD may fluctuate over time \cite{pahwa2010early}.

\begin{figure}[t]
\centerline{\includegraphics[height=32.3mm, width=77.90mm]{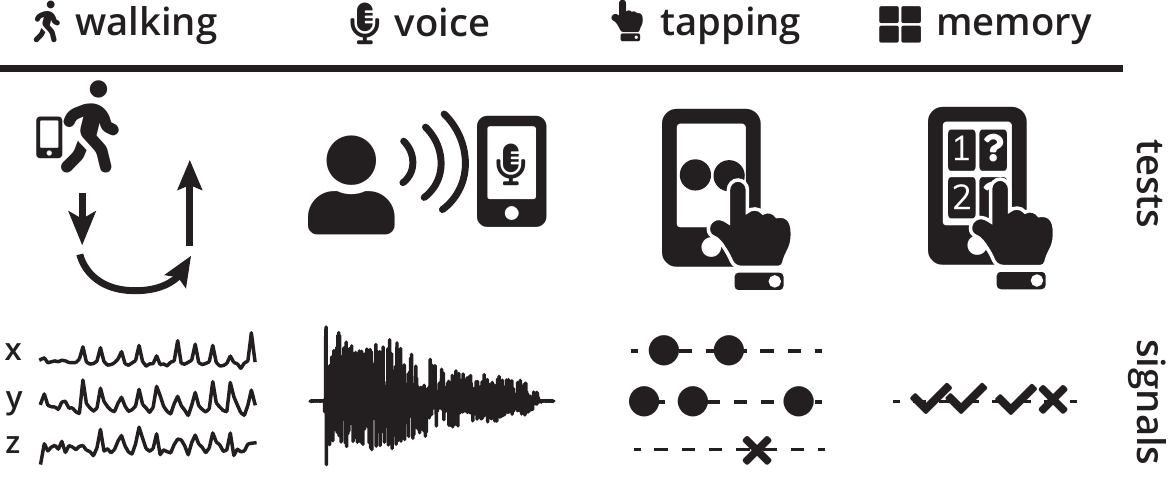}}
\caption{Smartphones can be used to perform tests that are designed to trigger symptoms of Parkinson's disease (top). During these tests, smartphone sensors record high-resolution signals (bottom) that we can use to distinguish between people with and without Parkinson's disease.} 
\label{fig:overview_tasks}
\end{figure}
Smartphone-based tests could potentially give clinicians access to long-term measurements of symptom severity and symptom fluctuation by enabling patients to record themselves outside the clinic (Figure \ref{fig:overview_tasks}). However, making sense of observational smartphone data is extremely challenging for both humans and machines due to the large number of diverse data streams sampled at high resolution over long periods of time. Major unsolved questions include how to simultaneously cover the wide range of symptoms associated with PD, how to best aggregate the vast amounts of clinically relevant data collected over time, and how to communicate the decisions of predictive models to clinicians. 

To address these issues, we present a novel approach towards distinguishing between people with and without PD from smartphone data. Our method is built on the idea of first training specialised models to assess symptom severity from single test instances, and then using an evidence aggregation model to aggregate an arbitrary number of assessments from several types of tests into a final prediction. We extend our method with hierarchical attention to visualise both the importance of tests as well as the importance of segments in those tests towards a prediction. Our experiments demonstrate that this approach leads to significant improvements in predictive performance over several strong baselines, and highlight the potential of smartphones to become accessible tools for gathering clinically relevant data in the wild.

\noindent\textbf{Contributions.} Our contributions are as follows:
\begin{itemize}[noitemsep]
\item We present machine-learning models to assess symptoms of PD from signals recorded during smartphone-based walking, voice, tapping and memory tests.
\item We introduce an evidence aggregation model (EAM) to integrate arbitrary numbers of symptom assessments from multiple types of tests over long periods of time to produce a single diagnostic score.
\item We develop a hierarchical neural attention mechanism that quantifies the importance of both individual tests and segments within those tests towards the diagnostic score.
\item We perform experiments on real-world data collected from 1853 mPower participants with and without PD that show that our approach leads to significant improvements in prediction performance over several strong baselines.
\end{itemize}

\section{Related Work}
\paragraph{Background.} Machine learning has a rich history in facilitating medical diagnoses. Machine learning has, for example, been applied to diagnosing breast cancer from tumor features \cite{zheng2014breast}, cardiac arrhythmias and cardiac risk factors from smartphone-based heart rate sensors \cite{oresko2010wearable,schwab2017beat,ballinger2018deepheart}, skin cancer from clinical images \cite{esteva2017dermatologist}, depressed moods from information self-reported via smartphones \cite{suhara2017deepmood}, and a wide range of clinical diagnosis codes from electronic health records and lab test results \cite{lipton2015learning,choi2016doctor,razavian2016multi}. Predicting a person's disease status is difficult because there is a vast range of factors that may influence an individual's health. Wearable sensors and smart devices enable us to capture a number of these factors with minimal burden on users by passively and continuously tracking behaviors and environmental factors \cite{quisel2017collecting}. However, in contrast to clean, standardised benchmark datasets, observational data collected by wearable sensors and smart devices in the real-world is often difficult to integrate with existing machine-learning approaches. The difficulty of applying existing machine-learning methods to complex datasets has led to the development of specialised methods to deal with several of the idiosyncrasies of observational health data, such as missingness \cite{lipton2016directly,che2018recurrent}, long-term temporal dependencies \cite{choi2016doctor}, noise \cite{schwab2017beat}, heterogeneity \cite{libbrecht2015machine}, irregular sampling \cite{lipton2015learning}, sparsity \cite{lasko2013computational}, and multivariate input data \cite{ghassemi2015multivariate,schwab2018not}. However, adapting existing machine-learning methods to account for the idiosyncrasies of healthcare data remains an ongoing challenge \cite{ghassemi2018opportunities}. 

\paragraph{Monitoring and Diagnosis of PD.} There has been much interest in leveraging new technologies and data modalities to better diagnose and assess symptom severity in PD. There are a number of driving factors behind the interest in new approaches: Firstly, despite the severity of the disease, clinical PD diagnoses are currently relatively inaccurate. Diagnoses are particularly difficult in the earlier stages of the disease and in the presence of other disorders that may appear similar to PD \cite{rizzo2016accuracy}. Secondly, new technologies could lead to patients receiving their diagnoses earlier. An early diagnosis could potentially improve a patient's quality of life by giving them access to symptom-suppressing treatments \cite{global2002factors}. Lastly, both clinical trials for new pharmaceutical treatments and clinical decision-making require the ability to accurately diagnose and objectively assess symptoms of PD \cite{shulman2006subjective,dorsey2017first}. Previous works have for example used data from pen movements \cite{smith2007diagnosis}, wearables \cite{patel2009monitoring,klucken2013unbiased}, and speech features \cite{little2007exploiting,little2009suitability,tsanas2010accurate,tsanas2011nonlinear,tsanas2012novel} to objectively monitor or diagnose PD. A number of works have also proposed the use of smartphone sensors for continuously monitoring symptoms in PD \cite{hammerla2015pd,arora2015detecting,zhan2016high,zhan2018using,prince2018ensemblepd}. Recently, the PD Digital Biomarker DREAM challenge\footnote{\url{http://synapse.org/DigitalBiomarkerChallenge}} aimed to develop machine-learning models to diagnose PD from accelerometer data in a collaborative effort. \cite{emrani2017prognosis} proposed a multitask-learning framework to identify biomarkers that are predictive of  progression in PD. However, their approach did not integrate raw sensor data and could not handle missing input data. 

In contrast to existing works, we present the first machine-learning approach to distinguishing between people with and without PD that integrates information from sensor measurements of several types of smartphone-based tests over long periods of time. Our approach is able to simultaneously (i) assess single test instances and (ii) produce a unified diagnostic score. In addition, we introduce a hierarchical neural attention mechanism that enables us to reason about both the importance of specific tests as well as the importance of individual segments within those tests towards the final diagnostic score. Furthermore, we perform our experiments on data collected from 1853 mPower participants, the largest cohort used to validate a machine-learning approach to diagnosing PD from smartphone data to date.

\section{Methodology}
\paragraph{Overview.} We utilise data collected during the mPower study, a large-scale observational study about PD conducted entirely through a smartphone app \cite{bot2016mpower}. In the study, participants with and without PD are asked to perform four smartphone-based tests (walking, voice, tapping and memory; Figure \ref{fig:overview_tasks}) up to three times a day without any supervision. In addition to regularly performing the tests, participants provide their detailed demographic profile, including possible prior clinical diagnoses of PD, using self-reporting forms within the app\footnote{\url{https://parkinsonmpower.org/}}. The main idea of the presented approach is to connect the sensor data collected by the participants' smartphones with their prior professional diagnoses to train machine-learning models to learn to diagnose PD. 

\paragraph{Smartphone Tests.} mPower participants perform the following four types of tests using their personal smartphones: 
\begin{itemize}
\item[\raisebox{-.15\height}{\includegraphics[height=10pt]{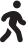}}]\textbf{Walking Test.} To perform the walking test, participants are asked to put their smartphone in their pocket, walk 20 steps forward, turn around, stand still for 30 seconds, and then walk 20 steps back. We denote the three distinct segments of the walking test as: Outbound, rest, and return, respectively. During the test, the smartphone's accelerometer and gyroscope record the participant's three-dimensional linear and angular acceleration. This test is designed to measure movement impairments associated with PD, such as tremor, rigidity, and freezing of gait.  

\begin{figure}[t]
\centerline{\includegraphics[height=36mm, width=75mm]{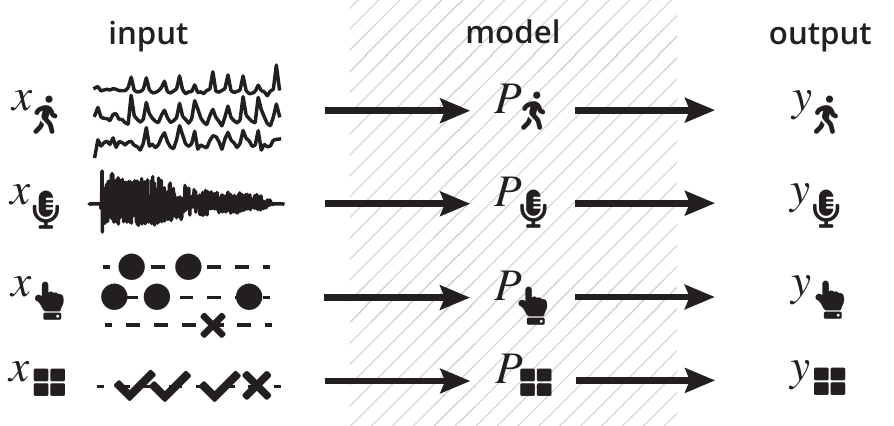}}
\caption{An illustration of the data processing pipelines for each of the test types (vertical) from the input signals $x_\star$ (left) over the specialised predictive models $P_\star$ (center) to the single-test output predictions $y_\star$ (right). The use of specialised predictive models for each test type enables us to choose the most suitable model for each of the heterogenous input signals.} 
\label{fig:tasks}
\end{figure}

\item[\raisebox{-.15\height}{\includegraphics[height=10pt]{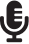}}]\textbf{Voice Test.} In the voice test, participants are asked to say "aaaah" into their smartphones' microphone for up to 10 seconds. The smartphone's microphone records the audio data during the test and during the preceding countdown. The goal of the audio test is to expose speech impairments that are commonly found in people with PD. 

\item[\raisebox{-.1\height}{\includegraphics[height=10pt]{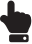}}]\textbf{Tapping Test.} In the tapping test, participants are asked to position their smartphones on a flat surface and alternatingly tap two buttons on the screen for 20 seconds. The smartphone records the positions and timestamps of the participant's taps on the screen. In addition, the smartphone's accelerometer measures the three-dimensional movements of the smartphone during the test. The tapping test is aimed at uncovering signs of impaired finger dexterity. Impaired finger dexterity is a common symptom in people with PD.

\item[\raisebox{-.15\height}{\includegraphics[height=8pt]{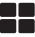}}]\textbf{Memory Test.} In the memory test, participants are presented with a grid of flowers on their smartphone screens. During the test, different flowers are illuminated one at a time. Participants are then asked to repeat the observed sequence by touching the flowers in the same order. The collected data includes the positions and timestamps of the participant's taps on the smartphone's screen and the sequence order as displayed to the participant. This test measures the spatial memory of the participant, which may be impaired due to PD \cite{bot2016mpower}.
\end{itemize}

\begin{figure}[t]
\centerline{\includegraphics[height=36mm, width=85mm]{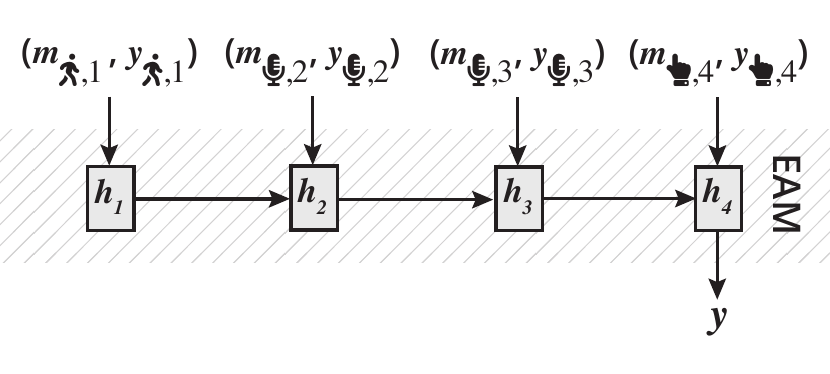}}
\caption{Temporal ensembling using an evidence aggregation model (EAM). The EAM (grey) receives per-test metadata ($m_{\star,j}$) and per-test output predictions ($y_{\star,j}$) in temporal order as input. In this example, the EAM's hidden state ($h_j$) aggregates the information from the $k=4$ performed tests to produce a final output $y$ that indicates whether or not the participant is likely to have PD.} 
\label{fig:temporal_ensemble}
\end{figure}

\paragraph{Multistage Approach.} Our approach to distinguishing between people with and without PD consists of two stages. In the first stage, we use specialised predictive models $P_\star$ to identify PD in signals $x_\star$ from a single type of test with $\star \in \{\raisebox{-.15\height}{\includegraphics[height=10pt]{img/walking}}\  \mbox{(walking)}, \  \raisebox{-.15\height}{\includegraphics[height=10pt]{img/voice}}\  \mbox{(voice)},\   \raisebox{-.15\height}{\includegraphics[height=10pt]{img/tapping}}\  \mbox{(tapping)},\   \raisebox{-.15\height}{\includegraphics[height=10pt]{img/memory}}\  \mbox{(memory)}\}$. The specialised models are trained to predict a participant's diagnosis given the signal data from exactly one sample instance of one type of test (Figure \ref{fig:tasks}). The output of the specialised models is a local prediction $y_\star$ that indicates, on a scale from 0 to 1, how likely it is that the participant that performed the given test instance has PD: 
\begin{align}
\label{eq:p_star} y_\star &= P_\star(x_\star)
\end{align}
The specialised models $P_\star$ are the building blocks for the second stage. In the second stage, the outputs $y_{\star,j}$, with $j \in [1, k]$, of the specialised models and the metadata $m_{\star,j}$ for all $k$ tests performed by a user are aggregated into a single diagnostic prediction $y$ using an EAM (Figure \ref{fig:temporal_ensemble}): 
\begin{align}
\label{eq:T} y &= \text{EAM}([(m_{\star,1}, y_{\star,1}), ..., (m_{\star,k}, y_{\star,k})])
\end{align}
\noindent The primary idea behind Equations \ref{eq:p_star} and \ref{eq:T} is to disentangle learning how to assess symptom severity from each test and how to aggregate multiple tests over a period of time. This compositional approach to modelling the problem of diagnosing PD from a range of diverse smartphone tests enables us to choose the most suitable predictive model for the various test types and the EAM. Furthermore, each specialised predictive model $P_\star$ is optimised for one type of test only. In contrast to an end-to-end model, the specialised predictive models do not need to consider how to aggregate multiple tests and which patterns may be important in other tests. Similarly, the EAM is entirely focused on learning how to best aggregate the evidence from multiple tests. In essence, our approach follows a divide-and-conquer approach by ensuring that each component is focused on exactly one task. Another benefit of the given abstract formulation is that it enables us to choose from a wide range of models for both the specialised predictive models and the EAM, since there are no specific requirements on either other than that they need to process $x_\star$ and tuples of $(m_{\star,i}, y_{\star,i})$, respectively.
{
\parfillskip=0pt
\parskip=0pt
\par} 

\begin{table*}[t]
\label{tb:results_tasks}
\begin{small}
\centering
\setlength{\tabcolsep}{1.24ex}
\begin{tabular}{lcc@{\hskip 3.15ex}@{\hskip 3.15ex}cc@{\hskip 3.15ex}@{\hskip 3.15ex}cc}
\toprule 
 & \multicolumn{2}{c}{\hskip -6.3ex \raisebox{-.22\height}{\includegraphics[height=10pt]{img/walking}}\hspace{0.5ex} outbound} &\multicolumn{2}{c}{\hskip -6.3ex \raisebox{-.22\height}{\includegraphics[height=10pt]{img/walking}}\hspace{0.5ex} rest} &\multicolumn{2}{c}{\hskip -3.15ex \raisebox{-.22\height}{\includegraphics[height=10pt]{img/walking}}\hspace{0.5ex} return}
\\
 & CNN & Feature & CNN & Feature & CNN & Feature \\ 
\midrule
AUC\hspace{6ex} & \textbf{0.53} (0.50, 0.56) & 0.50 (0.50, 0.53) & \textbf{0.53} (0.50, 0.56) & 0.52 (0.50, 0.55)  & \textbf{0.77} (0.74, 0.79) & \textbf{0.77} (0.75, 0.79) \\
AUPR & \textbf{0.60} (0.57, 0.64) & \textbf{0.60} (0.55, 0.62) & \textbf{0.62} (0.59, 0.66) & 0.61 (0.55, 0.62)  & 0.72 (0.53, 0.87) & \textbf{0.86} (0.84, 0.88) \\
\\ 
\toprule 
 & \multicolumn{2}{c}{\hskip -6.3ex \raisebox{-.22\height}{\includegraphics[height=10pt]{img/voice}}\hspace{0.5ex} voice} &\multicolumn{2}{c}{\hskip -6.3ex \raisebox{-.22\height}{\includegraphics[height=10pt]{img/tapping}}\hspace{0.5ex} tapping} &\multicolumn{2}{c}{\hskip -3.15ex \raisebox{-.15\height}{\includegraphics[height=10pt]{img/memory}}\hspace{0.5ex} memory} 
\\
 & CNN & Feature & CNN & Feature & RNN & Feature \\ 
\midrule
AUC\hspace{6ex}  & 0.53 (0.50, 0.55) & \textbf{0.56} (0.54, 0.58)  & 0.59 (0.57, 0.61) & \textbf{0.62} (0.60, 0.64) & \textbf{0.65} (0.60, 0.69) & 0.52 (0.50, 0.57) \\
AUPR & \textbf{0.48} (0.45, 0.51) & 0.45 (0.43, 0.48)  & 0.56 (0.53, 0.59) & \textbf{0.65} (0.62, 0.67) & \textbf{0.91} (0.88, 0.93) & 0.87 (0.84, 0.89) \\
\bottomrule
\end{tabular}
\end{small}
\caption{Comparison of the AUC and AUPR values for the different test types when only given the data of a single test to make a diagnostic decision. We compared the performances of neural networks (CNN, RNN) with expert features from biomedical literature fed to a random forest model (Feature) on the validation set. The listed models were the best models encountered over 35 hyperparameter optimisation runs for each test and model type. We calculated the 95\% confidence intervals (CIs) using bootstrap resampling with 1000 bootstrap samples. A comparison between the test types was not possible, because the evaluated subsets differed significantly due to different user groups preferring to do certain tests in different amounts (Appendix D). }
\end{table*}

\paragraph{Hierarchical Neural Attention.}
In addition to the diagnostic score $y$, our approach provides the clinician with information about which tests and test segments in the data recorded by the user were most important for the model's output. Presenting information about which data the model output is based on can help put the diagnostic score $y$ in perspective and inform the clinician's further clinical decision-making. For example, in a patient whose diagnostic prediction focused primarily on motor symptoms, the clinician can focus her efforts on ruling out other movement disorders that may cause similar symptoms. In order to highlight (i) which individual tests were most important for the EAM's output $y$, and (ii) which segments of specific tests were most important for the local predictions $y_\star$, we introduce a hierarchical neural soft attention mechanism. When using neural networks as predictive models, the upper-level attention mechanism (i) is a component of the EAM and the lower-level attention mechanism (ii) is part of the specialised models $P_\star$. Both the upper- and lower-level attention mechanism use the same mathematical formulation. Given the top-most hidden feature representations $h_i$ of (i) all the tests performed by a user, or (ii) segments in the recorded signal streams for a single test, we calculate attention factors $a_i$ using \cite{xu2015show,schwab2017beat,schwab2018granger}:
\begin{align}
\label{eq:a_i}
a_{i} &= \frac{\exp(u_{i}^Tu_{\text{s}})}{\sum_{j=1}^m \exp(u_{j}^Tu_{\text{s}})}
\end{align}
where
\begin{align}
\label{eq:u_i}
u_{i} &= \text{activation}(W_{\text{s}}h_{i} + b_{\text{s}})
\end{align}
Equation (\ref{eq:u_i}) corresponds to a single-layer MLP with a weight matrix $W_{\text{s}}$ and bias $b_{\text{s}}$. The single-layer MLP projects $h_i$ into a suitable hidden representation $u_i$ for comparison with $u_{\text{s}}$. We then calculate the attention factors $a_i$ by computing the softmax similarity of $u_i$ to  $u_{\text{s}}$. $u_{\text{s}}$ is the most informative hidden representation, i.e. the hidden representation for which $a_i$ would be the highest \cite{schwab2018granger}. $W_{\text{s}}$, $b_{\text{s}}$ and $u_{\text{s}}$ are learned parameters and jointly optimised with the other parameters during training.

\section{Experiments}
Our experiments aimed to answer the following questions:
\begin{enumerate}[noitemsep]
\item[1] What is the comparative performance of various specialised models $P_\star$ in diagnosing PD from a single test?
\item[2] How do EAMs compare to existing methods for aggregating multiple local predictions?
\item[3] What is the overall diagnostic accuracy of our approach?
\item[4] Does the proposed hierarchical neural attention mechanism identify meaningful data points?
\end{enumerate}
To answer these questions, we performed experimental comparisons between various baselines, predictive models and EAMs both on predicting PD from a single test and from an arbitrary number of tests. 

\paragraph{Dataset and Study Cohort.} We use data from the mPower study, a worldwide observational study about PD conducted entirely through smartphones \cite{bot2016mpower}. Starting in March 2015, the study recruited participants aged 18 and older around the world through a mobile app. Participants provided their demographic profile, including prior diagnoses of PD, through self-reporting, and performed the four test types regularly. Out of the study cohort, we used the subset of participants that were 45 or older, because there were very few participants in the dataset that had a clinical diagnosis at younger age. We used only those tests that were performed off medication, except for the memory tests. We performed a random split stratified by participant age to divide the available dataset into a training set (70\%), validation set (10\%), and test set (20\%). Each participant and the tests they performed were assigned to exactly one of the three folds without any overlap (Table \ref{tb:dataset}).

\paragraph{Models.} For each test type, we trained several specialised predictive models $P_\star$ using both automated feature extraction with neural networks and random forest (RF) models. We used expert features from biomedical literature that have been shown to be predictive of PD in the given data modalities as inputs to the RF models. The complete list of features \begin{table*}[!htpb]
\setlength{\tabcolsep}{1.24ex}
\label{tb:results_all}
\centering
\begin{small}
\begin{tabular}{lrrrr}
\toprule
Method & AUC & AUPR & F1 & Sens@95\%Spec\\
\midrule
EAM (Both) + age + gender & \hspace{2ex}\textbf{0.85} (0.81, 0.89) & \hspace{2ex}\textbf{0.87} (0.82, 0.91) & \hspace{2ex}0.81 (0.75, 0.85) & \hspace{2ex}\textbf{0.43} (0.19, 0.54) \\
EAM (Neural networks) + age + gender\hspace{8.5ex} & 0.84 (0.80, 0.88) & 0.86 (0.81, 0.90) & \textbf{0.82} (0.74, 0.86) & 0.33 (0.21, 0.51) \\ 
EAM (Feature) + age + gender & 0.84 (0.79, 0.88) & 0.86 (0.81, 0.90) & 0.76 (0.73, 0.84) & 0.40 (0.23, 0.56) \\
End-to-end neural network + age + gender & 0.50 (0.50, 0.56) & 0.54 (0.46, 0.62) & 0.27 (0.20, 0.70) & 0.04 (0.01, 0.07)  \\[1ex]
Age + gender & 0.74 (0.69, 0.79) & 0.75 (0.68, 0.82) & 0.72 (0.67, 0.79) & 0.16 (0.09, 0.31) \\[1ex]
EAM (Both) & 0.70 (0.64, 0.75) & 0.74 (0.66, 0.79) & 0.67 (0.60, 0.71) & 0.23 (0.15, 0.41) \\
EAM (Neural networks) & 0.71 (0.65, 0.76) & 0.75 (0.67, 0.80) & 0.67 (0.61, 0.72) & 0.24 (0.14, 0.41) \\
EAM (Feature) & 0.71 (0.66, 0.76) & 0.75 (0.67, 0.80) & 0.68 (0.61, 0.73) & 0.24 (0.14, 0.39) \\[1ex]
Mean Aggregation (Neural networks) & 0.64 (0.58, 0.69) & 0.67 (0.58, 0.73) & 0.60 (0.52, 0.68) & 0.22 (0.10, 0.27) \\
Mean Aggregation (Feature) & 0.62 (0.56, 0.68) & 0.60 (0.51, 0.66) & 0.62 (0.53, 0.69) & 0.13 (0.00, 0.19) \\
Max Aggregation (Neural networks) & 0.61 (0.55, 0.67) & 0.61 (0.53, 0.68) & 0.59 (0.54, 0.68) & 0.03 (0.01, 0.19) \\
Max Aggregation (Feature) & 0.61 (0.54, 0.66) & 0.61 (0.52, 0.68) & 0.60 (0.52, 0.65) & 0.07 (0.03, 0.18) \\
\bottomrule
\end{tabular}
\end{small}
\caption{Comparison of the AUC, AUPR, F1, and sensitivity at a fixed specificity of 95\% (Sens@95\%Spec) on the test set of 347 participants across the  methods that we evaluated. In parentheses are the 95\% CIs calculated with 1000 bootstrap samples.}
\end{table*}
\begin{table}[b!]
\setlength{\tabcolsep}{1ex}
\label{tb:dataset}
\centering
\begin{small}
\begin{tabular}{lrrrr}
\toprule
Property & Training & Validation & Test \\
\midrule
Subjects (\#) & 1314 (70\%) & 192 (10\%) & 347 (20\%) \\
PD (\%) & 52.36 & 50.00 & 56.20 \\
Female (\%) & 28.00 & 36.98 & 25.94 \\
Age (years) & 59.29 $\pm$ \hspace{1ex}9.40 & 59.53 $\pm$ \hspace{1ex}9.03 & 58.90 $\pm$ \hspace{1ex}9.24 \\
Walking (\#) & 13.89 $\pm$ 35.07 & 15.58 $\pm$ 33.90 & 14.03 $\pm$ 45.20 \\
Voice (\#) & 16.11 $\pm$ 40.21 & 19.47 $\pm$ 44.55 & 14.88 $\pm$ 45.12 \\
Tapping (\#) & 15.20 $\pm$ 38.04 & 18.50 $\pm$ 43.12 & 14.78 $\pm$ 42.67 \\
Memory (\#) & 14.01 $\pm$ 33.30 & 20.78 $\pm$ 35.92 & 17.58 $\pm$ 38.11 \\
Usage (days) & 24.27 $\pm$ 41.01 & 29.66 $\pm$ 45.73 & 25.43 $\pm$ 43.24 \\
\bottomrule
\end{tabular}
\end{small}
\caption{Population statistics of the training, validation, and test set. Numbers (\#) shown are mean $\pm$ standard deviation.}
\end{table}
used for each test type can be found in Appendix A. For the neural networks, we used different architectures of neural networks for each test depending on the type of input signal. For the walking, voice and tapping task, we used multi-layer convolutional neural networks (CNNs) with max pooling and temporal convolutions. For the memory test, we used a recurrent neural network (RNN) with bidirectional long short-term memory (BLSTM). Except for the voice test, the neural networks hosted the segment-level neural attention mechanisms described previously. For the voice CNN, we did not employ a neural attention mechanism because we found that it was detrimental to predictive performance. To implement the previously described EAM, we used a RNN architecture consisting of BLSTM cells. We trained EAMs using (1) only the RF models, (2) only the neural networks, and (3) an ensemble of both as specialised models to compare the performances of both approaches and whether their outputs are complementary. The detailed architectures for the neural networks and EAM are given in Appendix B. The EAM received a one-hot encoded unique identifier of the specialised predictive model as input metadata $m_{\star,j}$ with each local per-test prediction $y_\star$. The unique identifier enabled the EAM to differentiate between the various specialised predictive models. We additionally tested passing timing information, including the time since the last performed test and the hour of day at which the test was performed, for each performed test. However, we found no performance benefit in adding timing information to the metadata. Lastly, in order to determine whether the use of an EAM improves performance over simpler approaches, we  evaluated the performances of aggregating over local predictions $y_\star$ using the mean and maximum values of all local predictions. As a simple baseline based on demographics, we trained a MLP that received as input the age and gender of a participant and no information of any of their performed tests. To determine whether the proposed separation of learning to assess single test instances and learning to integrate multiple tests tests is beneficial, we also trained an end-to-end neural network jointly on both tasks. The end-to-end neural network used the same architectures as the specialised models to assess the tests and the same architecture as the EAM to integrate multiple tests. 

\paragraph{Hyperparameters.} We took a systematic approach to hyperparameter search. To avoid biases stemming from using different degrees of hyperparameter optimisation, we evaluated exactly 35 hyperparameter configurations for each trained specialised predictive model and EAM. We report the performances of those models which achieved the best validation set performances across the 35 runs. We selected the hyperparameters at random from a fixed range for each hyperparameter run. For the RF models, we used 512 to 1024 trees in the forest and a maximum tree depth between 3 and 5. For all neural networks, we used dropout of 0 to 70\% between hidden layers, an $L2$ penalty of 0, 0.0001 or 0.00001, and varying numbers of layers and hidden units depending on the test type (Appendix C). For the EAM, we used 2 to 5 stacked BLSTM layers with 16 to 64 hidden units each. We optimised the neural networks' binary cross-entropy for up to 500 epochs with a learning rate of 0.0001, a batch size of 32, and an early stopping patience of 12 epochs on the validation set. For memory reasons, we used a batch size of 2 for the end-to-end trained neural network. All other hyperparameters and hyperparameter ranges were exactly the same as in the separated models.
{
\parfillskip=0pt
\parskip=0pt
\par} \FloatBarrier
\begin{figure*}[t]
\centerline{\includegraphics[height=31.1182mm]{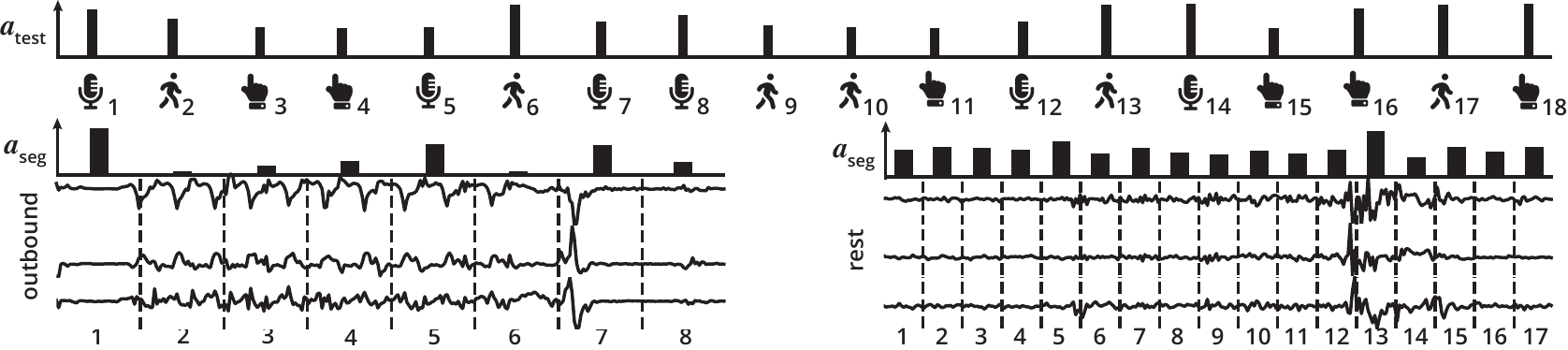}}
\caption{The outputs of the employed hierarchical neural attention mechanism on a user with PD that performed 18 tests. The timeline (top, left to right) shows all the tests performed by the user in temporal order. The tests performed (top, $a_{\text{test}}$) and the data segments within the tests (center, $a_{\text{seg}}$) were assigned attention weights that quantify their relative importance towards the final diagnostic score $y$. We show the outbound accelerometer data (left) and the rest accelerometer data (right) from walking test $\raisebox{-.15\height}{\includegraphics[height=10pt]{img/walking}}_{10}$. In the outbound recording, the attention mechanism focused strongly on the long delay to start moving (segment 1), increasingly choppy movement while setting foot (segments 3, 4, and 5), and the abrupt stop (segment 7). In the rest recording, we found that the attention was evenly distributed across the recording, likely because the whole recording contained signs of what could have been resting tremor. Sightly more attention was paid to segments with increased motion (segments 5 and 13). } 
\label{fig:atn} 
\end{figure*}

\paragraph{Preprocessing.} For computational performance reasons, we downsampled the input signals for the walking, voice and tapping test by factors of 10, 4, and 10, respectively.  In our initial evaluation, we did not see significant differences in predictive performance when using higher resolution data. After downsampling, we zero-padded or truncated the size of the sensor data to fixed lengths for each test type if they were too short or too long, respectively. The fixed lengths were 300, 250, 25, and 300 samples per record for the walking, voice, memory and tapping tests, respectively. For the voice test, we did not pass the raw voice signal to the neural networks. Instead, we passed the Mel-frequency cepstral coefficients (MFCC) that we extracted from the audio signal using a window size of 30 ms, a stride of 10 ms and 40 coefficients as input signal. For the RFs, we used the raw audio signals downsampled from their original sampling rate of 44100 Hz with factor 20 as inputs to the feature extractors. We standardised the accelerometer data for the walking and tapping tests to have zero mean and unit variance.

\paragraph{Metrics.} We computed the area under the receiver operating characteristic curve (AUC) and the area under the precision recall curve (AUPR) when comparing the different specialised predictive models. We evaluated the specialised models on the validation set to avoid test set leakage that could affect the evaluation of the models that aggregate information from multiple tests. We chose the best-performing specialised predictive models for each test for use in the aggregation models based on validation set performance. To compare the models that aggregated all available tests of a single user into a single diagnostic score, we additionally calculated the F1 score and the sensitivity at a fixed specificity level of 95\%. Some of the data folds were not balanced between people with and without PD. In particular, comparing single-test performances between test types was not possible due to differences in the number of tests performed between people with and without PD (Appendix D). We evaluated the performances of the three parts of the walking test (outbound, rest, and return) separately to determine their relative importances for diagnosing PD.

\section{Results}
\paragraph{Single-test Performance.} In terms of single-test performance, we found that, generally, both RFs with expert features and automated feature extraction with neural networks achieved competitive results for all tests (Table \ref{tb:results_tasks}). The performances of RFs with expert features and neural networks were similar across all tests, except for the tapping test, where the expert features significantly outperformed the neural networks, and the memory test, where the neural networks likewise outperformed the expert features. When comparing the three segments of the walking test, we found that return was the most informative for diagnosing PD.

\paragraph{Overall Performance.} We found large differences in performance between the various aggregation models that took into account all the performed tests of a user (Table \ref{tb:results_all}). Notably, EAMs outperformed all baselines by a large margin, and significantly improved upon the demographics-only model by integrating information from the tests performed by participants. We also found that expert features and neural network features were to some degree complementary, as the best EAM using both sets of predictive models outperformed its counterparts that only used one set of specialised predictive models. The neural networks trained end-to-end to simultaneously assess all types of tests and aggregate information from the available tests over time failed to converge. Closer analysis revealed that the end-to-end network was unable to effectively propagate gradients through the initially more attractive upper layers down to the per-test layers. Disentangling symptom assessment and temporal aggregation enabled EAMs to overcome this issue entirely.

\paragraph{Hierarchical Attention.} We plotted the attributions of the hierarchical neural attention mechanism against the raw signals of a sample participant with PD (Figure \ref{fig:atn}). In the walking tests, the attributions potentially corresponded to regions where signs of resting tremor and rigid motions could have appeared. In the memory tests, we found that the focus was directed at the difficult end stage of the test (Appendix E).

\section{Discussion}
Our work expands on prior studies \cite{arora2015detecting} by developing an effective methodology for integrating evidence from multiple types of smartphone-based tests over long periods of time, introducing tools to identify the most salient data segments across the vast amounts of generated data points, and evaluating these novel approaches in a large, representative cohort. The availability of smartphone-based tools for diagnosing PD could have a profound impact on clinical practice by enabling clinicians to access long-term observational data on patients. These additional data points could help give clinicians a more comprehensive and objective view on their patients' symptoms and symptom fluctuations, and therefore possibly enable more accurate diagnoses and treatment regimes. Another major potential benefit of enabling patients to record themselves with their smartphones is that it could enable clinicians to monitor their patients without requiring in-person visits that may be time-consuming and expensive, particularly in rural locations and developing countries. While our initial results are promising, further clinical validation is needed to determine whether the availability of smartphone data, the proposed diagnostic score, and in-depth information about the most relevant data points improve clinicians' ability to accurately diagnose PD. 

\paragraph{Limitations.} The main limitation of this work is that we use prior clinical diagnoses of users to train and evaluate our models. Clinical diagnoses for PD are themselves often inaccurate \cite{rizzo2016accuracy}, and are therefore not a flawless gold standard to evaluate against. In addition, much like in clinical assessments, smartphone-based tests depend on PD symptoms being clearly distinguishable for at least some of the tests being performed. While smartphones enable patients to record themselves when they believe that their symptoms are most pronounced, they still might not be clearly distinguishable against normal human variability, particularly in early-onset PD. Furthermore, the accuracy of smartphone diagnostics may be reduced when confronted with other movement and neurologic disorders that may appear similar to PD. More data, ideally from a prospective study, is needed to conclusively determine the robustness of machine-learning and smartphone-based tests against these confounding factors. 

\section{Conclusion}
We presented a machine-learning approach to distinguishing between people with and without PD from multiple smartphone-based tests. Our multistage approach is built on the idea of separately training (i) specialised models to assess symptom severity in instances of a single test, and (ii) an EAM to integrate all available single-test assessments into a final diagnostic score. In addition, we introduced a hierarchical attention mechanism that shows both which tests out of all performed tests, and which segments within those tests were most important for the model's decision. We demonstrated experimentally that the presented approach leads to significant improvements over several strong baselines with an AUC of 0.85 (95\% CI: 0.81, 0.89), an AUPR of 0.87 (95\% CI: 0.82, 0.91) and a sensitivity at 95\% specificity of 43\% (95\% CI: 0.19, 0.54) in data from a cohort of 1853 participants. Our results confirm that machine-learning algorithms and smartphone data collected in the wild over extended periods of time could in the future potentially be used as a digital biomarker for the diagnosis of PD.

\subsubsection*{Acknowledgments.}
This work was partially funded by the Swiss National Science Foundation (SNSF) projects 167302 and 150640. We acknowledge the support of NVIDIA Corporation with the donation of the GPUs used for this research. The data used in this manuscript were contributed by users of the Parkinson mPower mobile application as part of the mPower study developed by Sage Bionetworks and described in Synapse (doi:10.7303/syn4993293).

\small 
\bibliographystyle{aaai}
\bibliography{references.bib}



\section*{Supplementary material}

Table S4: Features used as inputs to the RFs used to assess tapping tests. The features were calculated on the inter-tap intervals and the button hit/miss time series. We additionally used the meta data associated with the memory test (overall score, number of games played, number of failures).

| Feature | memory | Reference / Brief description |
|---|---|---|
| Mean | | The mean of the amplitude of the input signal. |
| Standard deviation | | The standard deviation of the amplitude of the input signal. |
| Mean squared energy | | The mean squared energy of the amplitude of the input signal. |
| Mean Teager-Kaiser energy operator | | (Arora et al. 2015) |
| Cross-correlation | | The cross-correlation of the input signal with itself up to lag level 2. |
| Detrended fluctuation analysis | | (Arora et al. 2015) |
| Memory meta-data | | The meta-data of the memory test. |
| Fatigue$_{10\%}$ | | (Arora et al. 2015) |
| Fatigue$_{25\%}$ | | (Arora et al. 2015) |
| Fatigue$_{50\%}$ | | (Arora et al. 2015) |

Table S5: Population statistics of the training, validation, and test set for the walking tests. Numbers (#) shown are either absolute (Samples, Subjects), or mean ± standard deviation over all subjects.

| Property | Training | Validation | Test |
|---|---|---|---|
| Samples (#) | 8461 | 1496 | 2119 |
| PD (Samples, %) | 57.94 | 59.09 | 68.19 |
| Subjects (#) | 609 (71%) | 96 (11%) | 151 (18%) |
| PD (Subjects, %) | 43.19 | 44.79 | 47.68 |
| Female (%) | 26.77 | 34.38 | 19.21 |
| Age (years) | 59.71 ± 9.14 | 60.57 ± 8.56 | 58.62 ± 8.94 |
| Usage (days) | 31.08 ± 44.94 | 38.67 ± 51.52 | 32.50 ± 47.84 |

Table S6: Population statistics of the training, validation, and test set for the voice tests. Numbers (#) shown are either absolute (Samples, Subjects), or mean ± standard deviation over all subjects.

| Property | Training | Validation | Test |
|---|---|---|---|
| Samples (#) | 14176 | 2745 | 3586 |
| PD (Samples, %) | 54.36 | 49.25 | 69.77 |
| Subjects (#) | 880 (70%) | 141 (11%) | 241 (19%) |
| PD (Subjects, %) | 45.00 | 47.51 | 49.79 |
| Female (%) | 28.64 | 38.30 | 27.80 |
| Age (years) | 59.28 ± 9.29 | 59.65 ± 8.69 | 58.75 ± 9.36 |
| Usage (days) | 28.48 ± 43.72 | 33.42 ± 47.29 | 28.07 ± 44.88 |

Table S7: Population statistics of the training, validation, and test set for the tapping tests. Numbers (#) shown are either absolute (Samples, Subjects), or mean ± standard deviation over all subjects.

| Property | Training | Validation | Test |
|---|---|---|---|
| Samples (#) | 15823 | 2923 | 4064 |
| PD (Samples, %) | 51.85 | 48.58 | 66.63 |
| Subjects (#) | 1041 (70%) | 158 (11%) | 275 (19%) |
| PD (Subjects, %) | 42.84 | 42.41 | 48.00 |
| Female (%) | 28.05 | 37.34 | 27.27 |
| Age (years) | 58.59 ± 9.30 | 58.94 ± 8.75 | 58.35 ± 9.19 |
| Usage (days) | 25.16 ± 41.56 | 30.23 ± 45.83 | 26.48 ± 43.78 |

Table S8: Population statistics of the training, validation, and test set for the memory tests. We included memory tests done on medication, because there were few tests done off medication. Numbers (#) shown are either absolute (Samples, Subjects), or mean ± standard deviation over all subjects.

| Property | Training | Validation | Test |
| --- | ---: | ---: | ---: |
| Samples (#) | 4720 | 1143 | 1600 |
| PD (Samples, %) | 88.62 | 86.18 | 87.81 |
| Subjects (#) | 337 (70%) | 55 (11%) | 91 (19%) |
| PD (Subjects, %) | 64.09 | 65.45 | 75.82 |
| Female (%) | 35.01 | 27.27 | 28.57 |
| Age (years) | $61.11 \pm 9.05$ | $61.29 \pm 7.91$ | $61.93 \pm 8.87$ |
| Usage (days) | $56.34 \pm 57.19$ | $66.10 \pm 57.22$ | $59.32 \pm 60.65$ |

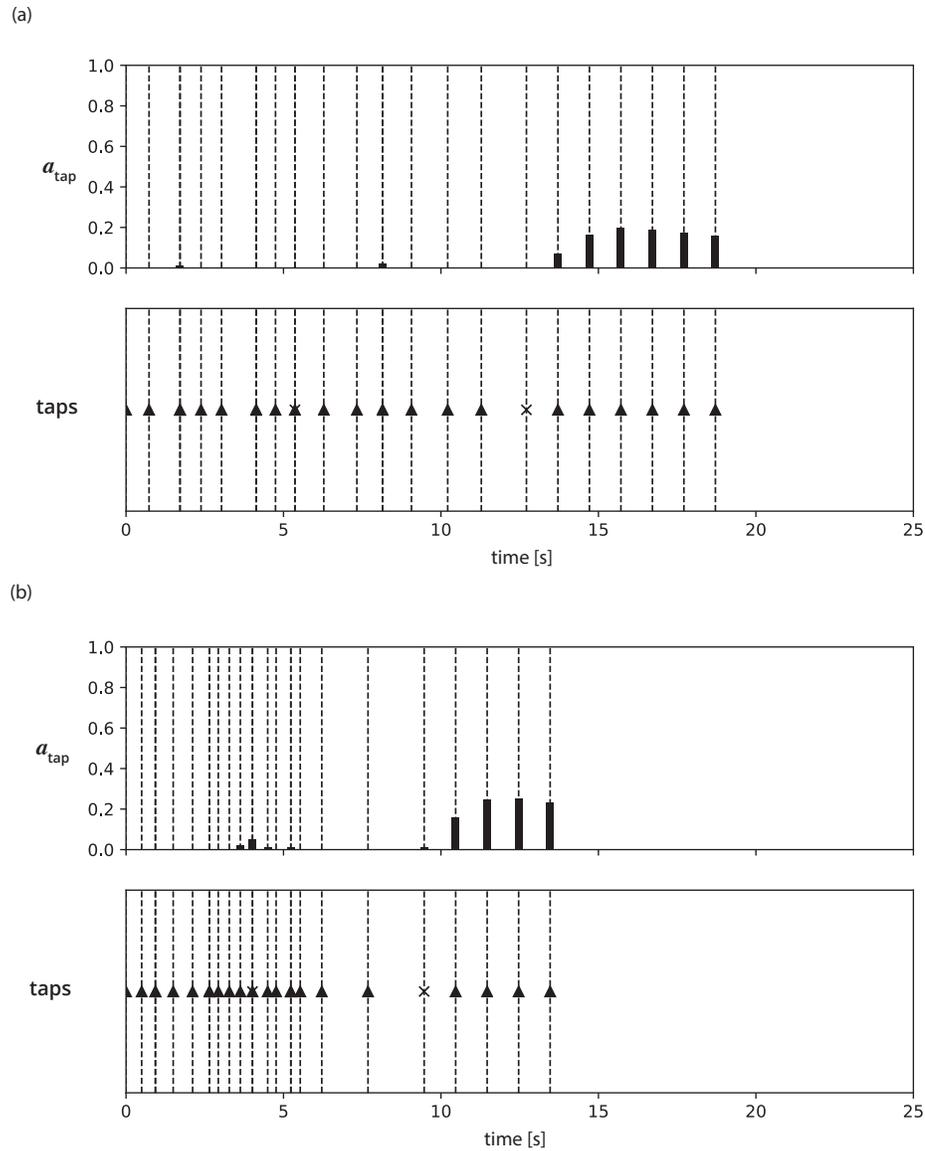

Figure S1: The outputs of the per-test neural attention mechanism ($a_{\text{tap}}$, top) on two representative samples (triangles and crosses, bottom) of memory tests from a user with Parkinson's disease. Triangles indicate correctly identified sequence elements and crosses indicate mistakes. We found that the predictive model's attention was typically focused on the more difficult final stage of the game. This pattern is visible in both samples (a) and (b). In both samples, we found that even mistakes in the early stage of the game do not receive a lot of attention relative to the more difficult end stage of the game.

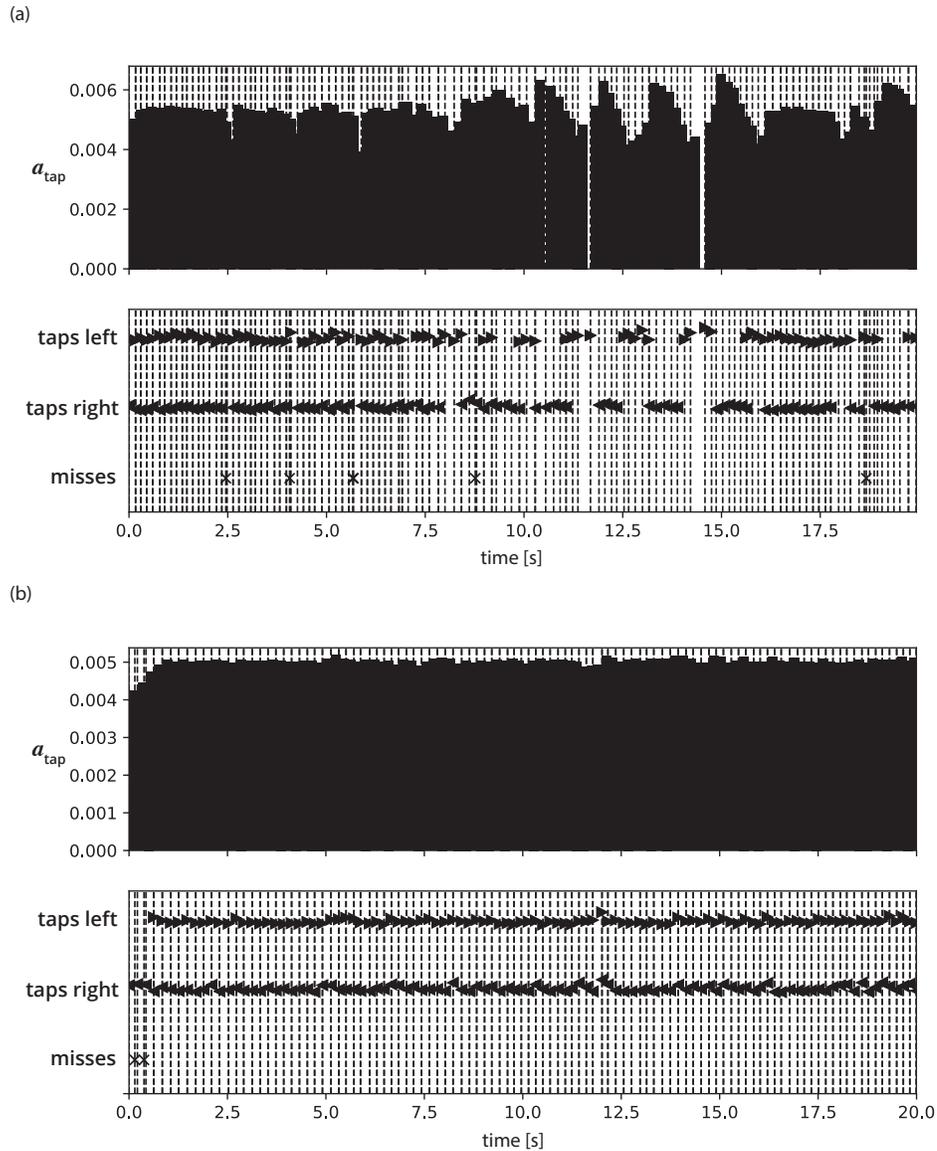

Figure S2: The outputs of the per-test neural attention mechanism ($a_{\text{tap}}$, top) on two representative samples (triangles and crosses, bottom) of tapping tests from a user with Parkinson's disease. The displacement of the tap triangles indicates the distance from the respective button's centre coordinate. In sample (a), we found that the assigned attention was typically lower in regions where the test was not performed properly (large breaks between taps, taps only on one side) and when taps were outside of the buttons' hit boxes (misses). In sample (b), we saw an almost uniform attention distribution, likely owing to the fact that the test was performed cleanly (aside from two misses at the start of the test). Our findings indicate that mistakes made in this test were not seen as predictive of Parkinson's disease. The predictive model instead focused on the user's overall tapping performance when the test was performed as intended. Furthermore, the predictive model distributed its attention among large numbers of taps, indicating that features spanning over many taps were in general seen as more predictive than individual taps.



\end{document}